\documentclass[aps,prd,superscriptaddress,
twocolumn,notitlepage,footinbib,bibnotes,floatfix,reprint,preprintnumbers]{revtex4-1}
\usepackage{graphicx}
\usepackage{epsf,amsmath,bbold,amsfonts,stmaryrd}
\usepackage{mathrsfs}
\usepackage{tikz}
\usetikzlibrary{arrows,matrix,positioning}
\usepackage{appendix}
\usepackage{amssymb}
\usepackage{float}
\usepackage{color} 
\usepackage{dsfont}
\usepackage{bm}
\usepackage[]{hyperref}
\hypersetup{pageanchor=false,citecolor=black,urlcolor=black}
\usepackage{indentfirst}
\usepackage[export]{adjustbox}
\usepackage{bigints}

\makeatletter

\def\a{\alpha}

\def\g{\gamma}

\def\D{\Delta}

\def\eps{\varepsilon}

\def\zb{\bar z}

\def\m{\mu}

\def\x{\xi}

\def\f{\frac}
\def\l{\left}

\def\p{\partial}
\def\r{\right}
\def\Tr{\text{tr}}

\newcommand{\av}[1]{\langle #1\rangle}

\newcommand{\tr}{{\rm tr }}
\newcommand{\bZ}{{\xoverline Z}}
\newcommand{\bX}{{\xoverline X}}
\newcommand{\bz}{{\bar z}}
\newcommand{\bx}{{\bar x}}

\def\be{\begin{equation}}
\def\ee{\end{equation}}

\def\bea{\begin{eqnarray}}
\def\eea{\end{eqnarray}}

\def\ba{\begin{array}}
\def\ea{\end{array}}

\def\bc{\begin{center}}
\def\ec{\end{center}}

\newsavebox\myboxA
\newsavebox\myboxB
\newlength\mylenA

\newcommand*\xoverline[2][0.75]{%
    \sbox{\myboxA}{$\m@th#2$}%
    \setbox\myboxB\null
    \ht\myboxB=\ht\myboxA%
    \dp\myboxB=\dp\myboxA%
    \wd\myboxB=#1\wd\myboxA
    \sbox\myboxB{$\m@th\overline{\copy\myboxB}$}
    \setlength\mylenA{\the\wd\myboxA}
    \addtolength\mylenA{-\the\wd\myboxB}%
    \ifdim\wd\myboxB<\wd\myboxA%
       \rlap{\hskip 0.5\mylenA\usebox\myboxB}{\usebox\myboxA}%
    \else
        \hskip -0.5\mylenA\rlap{\usebox\myboxA}{\hskip 0.5\mylenA\usebox\myboxB}%
    \fi}
\makeatother

\begin{document}
\title{Spontaneous Conformal Symmetry Breaking in Fishnet CFT }
\author{Georgios K. Karananas}
\email{georgios.karananas@physik.uni-muenchen.de}
\affiliation{Arnold Sommerfeld Center,
  Ludwig-Maximilians-Universit\"at M\"unchen,
         Theresienstra{\ss}e 37, 80333, M\"unchen, Germany}
      \author{Vladimir Kazakov}
      \email{vladimir.kazakov@phys.ens.fr}
      \affiliation{Laboratoire de Physique de l'\'Ecole
        Normale Sup\'erieure, 24 rue Lhomond, F-75231 Paris Cedex 05, 
        France }
      \affiliation{PSL Research University, CNRS, Sorbonne Universit\'e}
      \affiliation{CERN Theory Department, CH-1211 Geneva 23,
        Switzerland} 
\author{Mikhail Shaposhnikov}
\email{mikhail.shaposhnikov@epfl.ch}
\affiliation{Institute of Physics, 
Laboratory of Particle Physics and Cosmology,
\'Ecole Polytechnique F\'ed\'erale de Lausanne (EPFL),
CH-1015, Lausanne, Switzerland}

\preprint{CERN-TH-2019-147}

\begin{abstract}
  
Quantum field theories with exact but spontaneously broken conformal
invariance have an intriguing feature: their vacuum energy
(cosmological constant) is equal to zero.  Up to now, the only known
ultraviolet complete theories where conformal symmetry can be
spontaneously broken were associated with supersymmetry (SUSY), with
the most prominent example being the $\mathcal N$=4 SUSY
Yang-Mills. In this Letter we show that the recently proposed
conformal ``fishnet" theory supports at the classical level a rich set
of flat directions (moduli) along which conformal symmetry is
spontaneously broken.  We demonstrate that, at least perturbatively,
some of these vacua survive in the full quantum theory (in the planar
limit, at the leading order of $1/N_c$ expansion) without any fine
tuning. The vacuum energy is equal to zero along these flat
directions, providing the first non-SUSY example of a four-dimensional
quantum field theory with ``natural'' breaking of conformal symmetry.

\end{abstract}

\maketitle

\section*{Introduction}

Conformal Field Theories (CFTs) represent an indispensable tool to
address the behavior of many systems in the vicinity of the critical
points associated with phase transitions. They also describe the
limiting behavior of different quantum field theories deeply in the
ultraviolet (UV) and/or infrared (IR) domains of energy. Could it
be that CFTs are even more important and that the ultimate theory of
Nature is conformal?

At first sight, the answer to this question is negative. Indeed,
conformal invariance (CI) forbids the presence of any inherent
dimensionful parameters in the action of a CFT. Because of that, CFTs
have neither fundamental scales nor a well defined notion of particle
states.  On the other hand, Nature has both.

The loophole in these arguments is that conformal symmetry can be
exact, but broken spontaneously by the ground state. This breakdown
introduces an energy scale determined by the vacuum expectation value
of some scalar dimensionful operator. The notion of a particle is now
well defined, and in addition to massive excitations, the theory
contains a massless dilaton, the Goldstone mode of the broken CI.

Theories with spontaneous breaking of conformal symmetry may be
relevant for the solution of the most puzzling fine-tuning issues of
fundamental particle physics, namely the hierarchy and cosmological
constant problems. First, the Lagrangian of the Standard Model is
invariant under the full conformal group (at the classical level) if
the mass of the Higgs boson is put to zero. The observed smallness of
the Fermi scale in comparison with the Planck scale might be a
consequence of this~\cite{Wetterich:1983bi, Bardeen:1995kv}. Second,
if conformal symmetry is spontaneously broken, the energy of the
ground state is equal to zero (see,
e.g.~\cite{Amit:1984ri,Einhorn:1985wp,Rabinovici:1987tf,Shaposhnikov:2008xi}
and below). This fact may be relevant for the explanation of the
amazing smallness of the cosmological constant.

A systematic way to construct~{\em effective} field theories enjoying
exact but spontaneously broken CI was described
in~\cite{Shaposhnikov:2008xi}, following the ideas
of~\cite{Englert:1976ep,Wetterich:1987fm} (for further developments
see~\cite{Gretsch:2013ooa, Ghilencea:2015mza, Mooij:2018hew}, for a
review~\cite{Wetterich:2019qzx} and references therein). These
theories are free from conformal anomalies but
non-renormalizable. They remain in a weak coupling regime below the
scale induced by the spontaneous conformal symmetry breaking. Their
low energy limit may contain just the Standard Model fields, graviton
plus the dilaton, which essentially decouples and does not lead to a
long-range ``fifth'' force~\cite{Wetterich:1987fm,
Shaposhnikov:2008xb,Ferreira:2016kxi}. These theories are
phenomenologically viable and satisfy all possible experimental
constraints. Whether they can have a well-defined UV limit remains an
open question.

One can try to merge the ``bottom-up'' approach outlined above with
the ``top-down'' strategy, starting from a UV complete theory.  All
such known CFTs are~\emph{always} supersymmetric. The most notable and
well studied example is $\mathcal N=4$ SUSY Yang-Mills
(SYM). Although the immediate phenomenological relevance of such
theories is not clear, they are widely used as ``playgrounds'' for
studying the spontaneous breakdown of CI. 

In this~\emph{Letter} we show that there exists
a~\emph{nonsupersymmetric} CFT with these properties---the recently
proposed strongly $\g$-deformed $\mathcal N = 4$~SYM,
dubbed~\emph{Conformal Fishnet Theory}
(FCFT)\cite{Gurdogan:2015csr}~\footnote{The name of the theory stems
from the characteristic regular square lattice form of its planar
Feynman graphs}. This theory is well defined and finite at all scales
and has numerous flat directions at the classical level,~\emph{without
fine-tuning}. 

Moreover, some of them, are not lifted by  quantum
corrections, at least in the large-$N _ c$ limit~\footnote{To our
best knowledge, this is a unique behavior for a four-dimensional
theory, though a three-dimensional CFT with flat directions that
persist at the quantum level was presented
in~\cite{Rabinovici:1987tf}.}. 
We will be able to demonstrate this perturbatively in the coupling
constant. Among others, the reasons for these rather surprising
properties for a non-SUSY theory are:~\emph{i)}~its
UV-finiteness;~\emph{ii)} the fact that the FCFT has a large moduli
space, which increases the chances of finding directions along which
CI may be broken even without resorting to unnatural
tunings;~\emph{iii)}~the supersymmetric stabilization mechanism of the
parent theory is replaced by the absence of certain dangerous loop
diagrams that would normally lift the classical flat directions in the
Coleman-Weinberg (CW) effective potential~\cite{Coleman:1973jx}. This
self-protection mechanism is not powerful enough to completely
liberate the FCFT from all multiloop corrections on top of arbitrary
flat directions, even in the planar limit. In spite of that, only a
very limited sub-class of all higher loop graphs of $\phi^4$-type
theory (in the 't~Hooft limit) is present in the effective action. All
of them can be identified and their structure strongly hints towards
the presence of flat vacua which are robust under quantum effects.

Before moving on, let us emphasize that there is a price to pay for
these nice features: this chiral theory is not unitary. As a
consequence, it is a logarithmic
CFT~\cite{Joao-Caetano-unpublished-2017,Gromov:2017cja}. This is why
various parameters of the broken FCFT---e.g. the induced masses and
certain vertices on top of the flat vacua---are in general
imaginary. Nevertheless, the FCFT can be extremely useful as it
provides the so far unique possibility to test certain ideas of
potential phenomenological value in the non-SUSY world.

\section*{Fishnet CFT}

The FCFT involves the interacting $N_c \times N_c$ complex
matrix fields $X,\bX,Z,\bZ$ (if the theory were unitary a bar would
stand for Hermitian conjugation) in the adjoint of $SU(N_c)$; the
Lagrangian at the classical level
reads~\cite{Gurdogan:2015csr}~(see~\cite{Kazakov:2018hrh} 
for a review)
\be
\label{eq:lagr_biscalar}
\mathscr L = N_c\Tr \l( \p _ \m \bX \p _ \m X +\p_\m
\bZ \p _ \m Z  \\ 
+ \tilde\xi ^ 2 \bX \bZ X Z\r ) \ .
\ee
Here $\tilde \x = 4\pi
\x$, with the~\emph{real} coupling constant $\x$ defined as $\x ^ 2 =
g ^ 2N _ c \,e ^ {-i \g _ 3} /(4\pi)^2$; $g$ is the Yang-Mills
coupling constant and $\g_ 3$ one of the three twists of the parent
$\g$-deformed $\mathcal N=4$~SYM theory~\footnote{In this theory, the
$SO(6)$ R-symmetry is broken down to $U(1)^3$, with $\g_1,\g_2,\g_3$
being the parameters (twists) of the deformation.}. The
Lagrangian~(\ref{eq:lagr_biscalar}) is obtained by considering the
double-scaling limit corresponding to weak coupling and at the
same time large imaginary $\g _ 3$, such that $\x~\text{and}~\g _
{1,2}~\text{remain finite}$.

Let us briefly review the most general properties of FCFT in the
unbroken vacuum. A plethora of aspects of the theory on this conformal
phase have been and are still being investigated actively;
see~\cite{Gurdogan:2015csr,Gromov:2017cja,Chicherin:2017frs,Chicherin:2017cns,
Grabner:2017pgm,Kazakov:2018qez,Gromov:2018hut,Basso:2017jwq,
Basso:2018agi,Derkachov:2018rot,Korchemsky:2018hnb,Ipsen:2018fmu,
Basso:2018cvy,Kazakov:2018gcy,deMelloKoch:2019ywq,Gromov:2019aku,
Gromov:2019bsj,Chowdhury:2019hns}.

A direct consequence of the strong imaginary deformation is the
absence of the term corresponding to the Hermitian counterpart of the
quartic interaction. This makes manifest the fact that the theory is
not unitary.  On the other hand, it is exactly the absence of the
complex conjugate interaction term that has far reaching
implications. It restricts severely the number of possible planar
graphs for various physical quantities, to the point that, depending
on the physical quantity, there are often none, or only a handful of
diagrams, contributing at each order in the perturbative expansion.

At the same time, the fixed chirality of the interaction vertex, and
the absence of the vertex of opposite chirality, forces them to
possess the ``fishnet'' structure~\footnote{``Fishnet'' graphs
represent a regular square lattice of massless propagators with
vertices representing {$\phi^4$}-type interactions.}. This roughly
means that the bulk structure of sufficiently large planar graphs is
of the regular square lattice~\cite{Gurdogan:2015csr}.  Importantly,
the aforementioned chirality forbids the presence of certain diagrams,
such as the ones that induce masses for the fields and the ones that
renormalize the quartic coupling $\x$. Consequently, the FCFT behaves
as a fully-fledged logarithmic CFT, which implies the standard scaling
properties for its local observables (i.e. correlators).

In addition, the theory appears to be integrable in the planar,
't~Hooft $N_c\to\infty$
limit~\cite{Gurdogan:2015csr,Caetano:2016ydc,Gromov:2017cja}, due to
the integrability of the individual ``fishnet'' graphs discovered long
ago~\cite{Zamolodchikov:1980mb}, see
also~\cite{Chicherin:2012yn}~\footnote{It is not clear whether much of
this integrability stays intact in the spontaneously broken phase
considered throughout this paper; nevertheless, it can be certainly
useful in some particular calculations.}.  Hence, many of the physical
quantities---such as non-trivial Operator Product Expansion (OPE) data
as well as certain three- and four-point correlators---are in fact
exactly calculable~\cite{Gromov:2018hut}.

However, the model is not complete already at one-loop order: the
cancellation of the divergences associated with the correlation
functions of certain composite operators, such as
$\Tr(X^2),\Tr(\xoverline X^2), \Tr(\xoverline X Z),\Tr( X \xoverline
Z)$, requires that in the classical action~(\ref{eq:lagr_biscalar})
new double-trace terms be included~\cite{Fokken:2013aea}.  These read
\be
\label{eq:double-trace}
\begin{aligned}
\mathscr L _{d.t.}/(4\pi) ^ 2 = \a _ 1 ^ 2 \l [ \Tr(X^2)\Tr(\xoverline
X ^ 2) +\Tr(Z^2)\Tr(\xoverline Z ^ 2)\r ]& \\
-\a _ 2 ^ 2 \l [ \Tr
(XZ)\Tr(\xoverline X \xoverline Z) +\Tr(X\xoverline Z)\Tr(\xoverline X
Z) \r ]& \ ,
\end{aligned}
\ee
with $\a _ 1$ and $\a _ 2$ couplings that, in general, depend on the
renormalization scale, thus destroying, on the quantum level, the
conformal symmetry.  However, the beta functions for the running
double-trace couplings possess two complex conjugate fixed lines,
parametrized by $\xi$, with $\a_1^2=\pm \f {i\xi ^ 2 }{2} - \f {\xi ^
4 } {2} \mp \f {3i\xi ^ 6} {4} +\mathcal O (\xi ^ {8}) $ and $\a _ 2 ^
2 = \xi ^ 2$, for both of them~\cite{Sieg:2016vap,Grabner:2017pgm}.

The FCFT is completely defined by the explicitly local Lagrangian
$\mathscr L+\mathscr L _{d.t.}$, with conformal symmetry persisting at
the quantum level for the critical values of the $\a$'s.

\section*{Flat vacua}

\begin{figure*}[!t]
    \centering
    \includegraphics[scale=.45]{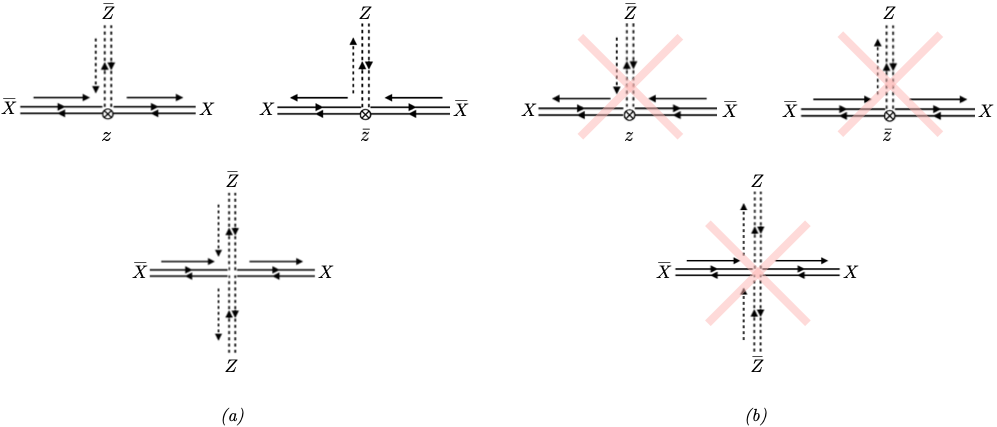}
    \caption{\emph{(a)}~Some of the cubic (upper graph) and the
      quartic (lower graph) interaction vertices that stem from the
      single-trace term when the Lagrangian is expanded around the
      symmetry-breaking vacua. A solid (dashed) line stands for the
      excitations $X$ or $\bX$ ($Z$ or $\bZ$), and ``$\pmb{\otimes}$'' for
      the vacuum expectation value $z$ or $\bz$.~\emph{(b)}~The cubic (upper
      graph) and quartic (lower graph) vertices of opposite chirality are
      absent in the FCFT. To highlight this fact, we have crossed out these
      nonexistent vertices.}
    \label{fig:vertices_full}
\end{figure*}

The spontaneous breaking of CI corresponds to a situation in which at
least one of the fields has a non-vanishing vacuum expectation value
(vev). As our CFT is non-unitary, we model this vacuum state by an
extremum of the (complex) effective action.

It is important to keep in mind that once we find such a (nontrivial)
saddle point, then the vacuum energy of the system automatically
vanishes along this flat direction. This follows from the fact that
for CFTs the potential $V$ is in general a homogeneous function of the
fields $\phi_i$ of the theory. In other words $V \propto \phi_i \f
{\partial V}{\partial \phi_i}$, where summation over repeated indices
is assumed. Provided that at least one of the fields acquires a
(constant in spacetime) vev, say $\hat \phi_1 \neq 0$, such that
\be
\f {\partial V}{\partial \phi_1}\Bigg\vert_{\hat \phi_1} = 0 \ ,
\ee
then it immediately follows that $V = 0$, although mass scale(s) are
now present in the theory.

Let us look for an ansatz that extremizes the potential of the
FCFT. To this end, we perform the following shifts in the action
\be
\label{eq:shifts}
X\to x+ X,~~~\bX\to \bx +\xoverline X,~~~Z\to z+Z,~~~\xoverline Z\to
\bz +\xoverline Z\ ,  
\ee
where $x,\bx,z,\bz$ are the vevs of the corresponding fields, and in
an abuse of notation we denoted the fluctuations again by
$X,\bX,Z,\bZ$ (as usual, these have zero vev's).

The matrix equations of motion are obtained by varying the effective
action w.r.t. $x,\bx,z,\bz$, respectively; they read
\begin{widetext}
\be
\begin{aligned}
\label{eq:gen_eq1}
 &-\kappa\langle\Tr(\bX^2)\rangle x\,
 -\langle\Tr\left(\bX\bZ\right)\rangle\,   \,  z-\langle\Tr\left(\bX
   Z\right)  \rangle\,   \bz+ N_c\,z\langle \bX\bZ\rangle
 + N_c\,\langle Z\bx\bZ\rangle+ N_c\,\langle
 Z\bX\rangle\bz+N_c\,\langle Z\bX\bZ\rangle=\\ 
&=\Tr(\bx^2) x+\Tr\left(\bx\bz\right)z+\Tr\left(\bx z\right) \,  \bz-
N_c\, z\bx\bz \ , 
\end{aligned}
\ee
\be
\begin{aligned}
\label{eq:gen_eq2}
&-\kappa\langle\Tr(X^2)\rangle \bx\,
-\langle\Tr\left(X\bZ\right)\rangle\,  \,   z-\langle\Tr\left(X
  Z\right)  \rangle\,  \bz+N_c\,\langle\bZ  X Z\rangle 
+ N_c\,\bz\langle XZ\rangle+ N_c\,\langle \bZ x Z\rangle+ N_c\,\langle
\bZ X\rangle z= 
\\
&=\Tr(x^2) \bx+\Tr\left(x\bz\right)z+\Tr\left(x z\right) \,  \bz-
N_c\, \bz x z \ , 
\end{aligned}
\ee
\be
\begin{aligned}
\label{eq:gen_eq3}
 & -\langle\Tr(\bX\bZ)\rangle x\,
 -\langle\Tr\left(X\bZ\right)\rangle\,
 \bx-\kappa\langle\Tr\left(\bZ^2\right)  \rangle\,  z+ N_c\,\bx\langle
 \bZ X\rangle 
+ N_c\,\langle \bX \bz X\rangle+ N_c\,\langle \bX\bZ \rangle x +N_c\,
\langle\bX  \bZ X\rangle=
\\
&=\Tr(\bx \bz) x+\Tr\left(x\bz\right)\bx+\kappa \Tr\left(\bz^2\right)
\,  z- N_c\, \bx  \bz x \ , 
\end{aligned}
\ee
\be
\begin{aligned}
\label{eq:gen_eq4}
  & -\langle\Tr(\bX Z)\rangle x\,  -\langle\Tr\left(XZ\right)\rangle\,
  \bx-\kappa\langle\Tr\left(Z^2\right)   \rangle\,  \bz+ N_c\,x\langle
  Z \bX \rangle 
 + N_c\,\langle Xz \bX \rangle+ N_c\,\langle XZ \rangle\bx+N_c\,
 \langle XZ\bX\rangle= 
\\
&=\Tr(\bx z) x+\Tr\left(x z\right)\bx+\kappa \Tr\left(z^2\right)  \,
\bz- N_c\, xz\bx  \ . 
\end{aligned}
\ee
\end{widetext}
Here $\kappa = -2 \a^2_1 /\x^2$, $\langle\dots\rangle$ denotes the
quantum average of the corresponding quantity w.r.t. the action
with the shifted fields~(\ref{eq:shifts}), and we took into account
the planar limit.

Notice that the presence of the non-Hermitian single-trace interaction
term, as well as the fact that $\kappa$ is complex at the conformal
point, results into the equations for the fields and their would-be
Hermitian counterparts to not be related by complex conjugation. In
turn, the solutions to the above equations for the pairs $x,\bx$ and
$z,\bz$ need not necessarily be complex conjugates, so the vev's may
be viewed as four independent constants in the space of matrix
fields. We will see that this may have important consequences for the
quantum fate of the flat directions.

\subsection*{Classical flat vacua}

Turning to the existence of (nontrivial) vacua, we note that
classically, i.e. in the tree approximation, all the deviations
$X,\bX,Z,\bZ$ of the fields in~(\ref{eq:gen_eq1})-(\ref{eq:gen_eq4})
should be put to zero (and there is no quantum average). Thus, the
classical flatness conditions are reduced to the r.h.s. of these
equations being zero.

For simplicity, we will work with configurations for which $x_{\rm
tree}= \bx_{\rm tree}=0$~\footnote{We can also relax the requirement
that {$x _ {\rm tree} = 0$} and require that both fields have nonzero
vev. This considerably enlarges the set of possible flat vacua. For
instance, field configurations such that {$x _ {\rm tree} \propto z _
{\rm tree}$}, may provide yet another set of acceptable vacua along
which CI is nonlinearly realized.}, such that the first two of the
equations of motion are identically satisfied, while the last two
become (since $\kappa\neq 0$)
\be
\label{eq:eom_tree}
 \Tr\l(\bz_{\rm tree}^2\r)  z_{\rm tree}=
 0 \ ,~~~\text{and}~~~\Tr\l(z_{\rm tree}^2\r) \bz_{\rm tree}=0 \ ,  
 \ee
with $z_{\rm tree}$ and $\bz_{\rm tree}$ (constant) classical fields 
subject to
\be
\label{eq:unimod_constr}
\Tr\left(z_{\rm tree}\right)= \Tr\left(\bz_{\rm tree}\right)=0 \ ,
\ee
due to the unimodularity of the global $SU(N_c)$ symmetry. Inspection
of~(\ref{eq:eom_tree}) reveals that, at least at the classical level,
the fishnet CFT has a plethora of nontrivial symmetry breaking
solutions, at any value of the coupling $\x$~\footnote{Additional flat
directions open up at isolated values of {$\x$}.}.~Interestingly, some
are not present in the full $\mathcal N = 4$~SYM nor in its
$\g$--deformed descendant; rather, they emerge when the strong
imaginary {$\g$}-deformation limit---leading to the fishnet CFT---is
considered. A complete classification of the moduli space of the FCFT,
however, lies well beyond the scope of the present paper. Therefore,
here we will focus on the simplest possible symmetry breaking flat
vacua that we have been able to find and leave the search and study
for more complicated ones for the future.

The first option is to take $z_{\rm tree}$ and $\bz_{\rm tree}$ to be
nonzero, related by complex conjugation, and diagonal~\footnote{The
measure of the functional integral (and the original unbroken action)
is invariant w.r.t. arbitrary complex matrix rotations
$(X,\bX,Z,\bZ)\to U^{-1} (X,\bX,Z,\bZ)U$.  Using it we can reduce, in
general, only one of the four vev matrices $(z,\bz,x,\bx)$ to diagonal
form.}, i.e.
\be
\begin{aligned}
\label{eq:Zvev}
  &z_{\rm tree} = v\, \text{diag}\left (z_1, \ldots,z_{N_c}  \right) \
  , \\
  &\bz_{\rm tree} = \bar v\, \text{diag}\left(\bz_1, \ldots,\bz_{N_c} 
    \right) \ , 
\end{aligned}
\ee
with $v$ a (complex) parameter with dimension of mass and $z _ k$ are,
in general, complex numbers~\footnote{For the diagonal ansatz,
condition~(\ref{eq:unimod_constr}) translates into $ \Sigma_ {k=1} ^
{N _ c} z _ k= \Sigma _ {k=1} ^ {N _ c}\bar z _ k= 0$. }.~Since by
construction $z_{\rm tree}\neq 0$ (and consequently $ \bz_{\rm
tree}\neq 0$), the only option for both equations to hold is to
require that
\be
\label{eq:constr_2}
\sum _ {k=1} ^ {N_c} z _ k ^ 2 = \sum _ {k=1} ^ {N_c} \bar z _ k ^ 2= 
0 \ .   
\ee

The second class of symmetry breaking solutions to
eqs.~(\ref{eq:gen_eq1})-(\ref{eq:gen_eq4}) comprises vacua for which
the fields $X$ and $\bX$, and $Z$ and $\bZ$, are not related by
complex conjugation. As we have already pointed out, this is certainly
a possibility, due to the non-Hermiticity of the theory. We may
therefore assume that $z_{\rm tree},\bz _{\rm tree}\neq 0$ and subject
to~(\ref{eq:eom_tree}) and~(\ref{eq:unimod_constr}). As we will show
in the next section, such configurations may be rather interesting
when it comes to quantum corrections.

Yet another acceptable choice is to put $x_{\rm tree} =\bx_{\rm tree}
=\bz _{\rm tree} = 0$, while $z_{\rm tree}$ can be an arbitrary
traceless $N_c \times N_c$ matrix. Interestingly, even though
conformal symmetry is broken spontaneously along such flat directions,
the spectrum of the theory contains only massless degrees of freedom,
at least in the planar limit.

The third and final category of ``natural'' flat directions we will be
reporting on here involves nilpotent matrices of index 2, i.e.~$ z _
{\rm tree}\neq 0$, while~$z _ {\rm tree} ^ 2 =0$. Interestingly, such
vacua appear also in beyond the Standard Model phenomenology,
see~\cite{Maiezza:2016ybz}. Like in the previous case---and unlike
what happens with Hermitian theories---all the excitations on top of
these vacua in the leading $N_c$ order are massless, in spite of the
fact that conformal symmetry is clearly broken. More details on the
spectrum of excitations around the aforementioned classes of vacua can
be found in the Appendix~\ref{sec:appendix}.

Before moving on, let us stress that the existence of flat directions
for arbitrary $\x$---even at the classical level---is a rather salient
point that deserves some discussion.  One might expect that whether or
not the theory possesses ground states with nonlinearly realized
conformal symmetry would crucially depend on the specific value of the
coupling constant. This is precisely what happens in other
nonsupersymmetric CFTs such as the massless $\phi ^ 4$ theory and its
generalizations~\cite{Shaposhnikov:2008xb}, where finetunings are
required in order for CI to be spontaneously broken down to
Poincar\'e~\cite{Fubini:1976jm}, see
also~\cite{Coradeschi:2013gda}. On the contrary, the FCFT has many
vacua (some of which are inherited from its parent $\mathcal N =
4$~SYM) with vanishing energy, without the need for
finetuning. Equivalently, the dilaton---that is part of the theory's
spectrum in the Coulomb phase---has zero mass, naturally.

In the following we will argue that this phenomenon persists at the
quantum level and in the planar limit, at least for some of the vacua
we found.

\subsection*{Quantum Coleman-Weinberg effective potential}

To study the fate of conformal symmetry breaking at the quantum level,
we will also confine ourselves to ``$z$-vacua,'' for which
$x=\bx=0$. It is important to keep in mind that, with such an ansatz,
the extrema of the effective action do not break the discrete symmetry
$X\to -X,\, \bX\to -\bX$, meaning that we can drop all terms
containing averages with odd powers of these two fields
from~(\ref{eq:gen_eq1})-(\ref{eq:gen_eq4}).

 Consequently, only the last two of these equations survive and boil
down to~\footnote{Note that \(z,\bz\) are still arbitrary matrices, so
that the order should be respected.}
\be
\begin{aligned}
\label{1loop_tadpoles}
 &\tfrac{\kappa}{N_c}\Tr\l(\bz^2\r)z =
\l\langle \bX \bz X\r\rangle-\tfrac{\kappa}{N_c}\l\langle\Tr(\bZ^2) \r
\rangle z+  \l\langle\bX  \bZ X\r\rangle,\\
 &\tfrac{\kappa}{N_c}\Tr\l(z^2\r)\bz=
\l\langle X z \bX\r\rangle-\tfrac{\kappa}{N_c}\l\langle\Tr(Z^2) \r
\rangle \bz+  \l\langle X  Z \bX\r\rangle. 
\end{aligned}
\ee

\subsubsection*{The one-loop effective potential}

Whether or not quantum corrections jeopardize the CI by uplifting the
flat directions can be demonstrated already at the first loop order, by
investigating the CW effective potential~\cite{Coleman:1973jx}.

\begin{figure}[!h]
    \centering
    \includegraphics[scale=.3]{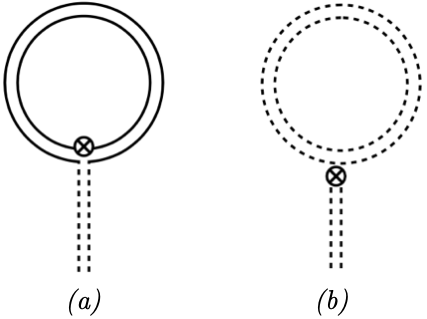}
    \caption{The one-loop tadpole diagrams stemming from the single-
      and double- trace terms---$(a)$ and $(b)$, respectively.  A solid
      (dashed) line stands for the excitations $X$ or $\bX$ ($Z$ or $\bZ$),
      and ``$\pmb{\otimes}$'' for the vacuum expectation value $z$ or
      $\bz$. We have suppressed the color and chirality arrows.}
    \label{fig:1loop_tadpoles}
\end{figure}

In this approximation $\a_1^2 = \pm i\xi^2/2$, so we can set
$\kappa=\pm i$. In addition, the last terms in~(\ref{1loop_tadpoles})
are irrelevant (they correspond to higher order Feynman graphs). In
the planar limit, the second and third terms are given by the diagrams
presented in Fig.~\ref{fig:1loop_tadpoles}.

As we did in the classical considerations, let us require that $z$ and
$\bz$ be diagonal matrices. The evaluation of the one-loop tadpole
diagrams is straightforward in this case, see
Appendix~\ref{sec:appendix2}. The matrix
equations~(\ref{1loop_tadpoles}) take the explicit form
\begin{align}
\label{eq:1loop_CW}
&\pm i\Tr\left(\bz^2\right)= \xi^2
 \left[\Tr(\bz^2)\log\frac{z}{\sqrt{Q}} +\Tr\left( \bz^2\log\frac{ \bz
                      }{\sqrt{Q}}\right) \right],& 
\nonumber\\[-5pt]
& &\\[-5pt]
&\pm i\Tr\left(z^2\right) =  \xi^2
 \left[\Tr(z^2)\log\frac{\bz}{\sqrt{Q}} +\Tr\left( z^2\log\frac{z
              }{\sqrt{Q}}\right)\right],\nonumber& 
\end{align}
with $Q =\sqrt{\tr(z^2)\,\tr(\bz^2)}$; note that due to the
non-Hermiticity of the FCFT the sign in the l.h.s. of both equations
must be the same---either plus or minus. The absence of sources
breaking explicitly the CI of the theory translates into the effective
potential (and its derivatives) to exhibit no dependence on the 't
Hooft-Veltman renormalization scale $\mu$. In turn, the derivatives of
the potential w.r.t. the fields are related to the beta function of
$\xi$, which vanishes by construction. Let us stress that at large
$N_c$, none of the physical quantities---such as correlators of local
fields---can actually depend on $\mu$ for the chosen background fields
$z,\bz$, since in the UV regime the theory behaves like in the
unbroken phase, which is UV finite. The CW potential is yet another
example of such a quantity.

We notice immediately that it is in principle possible to put to zero
the tree-level and one-loop parts of the potential separately,
provided that the vacuum~(\ref{eq:Zvev}), apart from the
constraints~(\ref{eq:unimod_constr}) and~(\ref{eq:constr_2}), is also
subject to
\be
\label{eq:constr_3}
\sum _ {k=1} ^ {N _ c} z _ k ^ 2 \log z _ k = \sum _ {k=1} ^ {N _ c}
\bar z _ k ^ 2 \log \bar z _ k  = 0 \ . 
\ee

This condition picks up a particular subclass of the classical vacua
discussed in the previous section. At the one-loop order these are not
lifted by quantum effects.  As a result, the vacuum energy of the loop
corrected theory on top of these flat directions is zero, or in other
words, the masslessness of the dilaton persists at one-loop level. It
should be stressed that this is a~\emph{unique} situation for a non-SUSY
four-dimensional theory.

Let us give a simple example of a flat vacuum which is robust under
one loop quantum corrections. Take $x=\bx=0$ and $z$ to be a
block-diagonal matrix comprising $N _ c/4$ diagonal sub-blocks each
with dimensions $4\times 4$
\be
\label{eq:block-mat-Z}
z= v\, \text{diag}(\underbrace{z_1,z_2,z_3,z_4},
\underbrace{z_1,z_2,z_3,z_4},\ldots) \ ,
\ee 
and $\bz$ is its Hermitian conjugate in this
case. Plugging~\eqref{eq:block-mat-Z} into the system of
transcendental eqs.~(\ref{eq:unimod_constr}),~(\ref{eq:constr_2})
and~(\ref{eq:constr_3}), we numerically find a complex (as a
consequence of the non-unitarity) solution
\begin{eqnarray}
\label{eq:explicit_su4}
&z_1& = -0.587849-0.808971\,i\,,~z_2 =0.260305+  1.45187\,i
      \,,\nonumber\\ &z_3& =1.32754- 0.642903\,i\, ,~~~~\,\, 
z_4 =-1 \, ,
\end{eqnarray}
where the overall rescaling \(z_j\to {\rm const}\times z_j\) was
absorbed into the complex modulus \(v\) labeling the one-parameter
family of flat vacua~\footnote{The masses generated on top of this
vacuum can be calculated from the quadratic variation of the full
effective potential \(V_{\rm eff}\) w.r.t. matrix fields \(Z,\bar
Z,X,\bar X \). The spectrum of the theory in the leading order at this
limit comprises: \emph{i)}~$N _ c^2 - 1$ complex massive excitations
of the matrix scalar $X$ whose masses are proportional to $\tilde \xi
^ 2 |v|^2\bar z _ i z _j$ with $z's$
from~(\ref{eq:block-mat-Z}),~(\ref{eq:explicit_su4}); \emph{ii)}~$N _
c^2-1$ gapless modes---including the dilaton which is proportional to
$\Tr\l(\bz Z +z \bZ\r)$. Note that beyond the planar approximation,
the excitations of $Z$ will acquire masses, as follows from the
variation of the CW action.}.

There is no difficulty in finding more examples for larger block
matrices of the form~(\ref{eq:block-mat-Z}), and thus with more of the
parameters labeling the flat vacua. For instance if we solve the
system of eqs.~(\ref{eq:unimod_constr}),~(\ref{eq:constr_2})
and~(\ref{eq:constr_3}) for $z$ made of $N_c/5$ sub-blocks of
dimensions $5\times 5$, we will have an extra parameter, in addition
to \(v\), parametrizing the flat directions. We can also mix
sub-blocks of different sizes.

Although this is certainly an interesting option, as we will now
demonstrate, it is not the only one. Actually, it is possible to
arrange a situation in which the tree-level and one-loop contributions
are of the same order of magnitude and can in principle balance each
other out. Remarkably, this enables the perturbative analysis of the
flat vacua and is in close analogy to what happens in the CW effective
potential in gauge theories~\cite{Coleman:1973jx}. In the present
context, we can achieve this by keeping the order of magnitude of the
vacuum fields $z_k,\bz_k\sim 1$, while $\tr (z^2)\sim\tr (\bz^2) \sim
N_c\,{\cal O}(\xi^2)$.  To this end, let us stick to vacua comprising
diagonal matrices, assume that $z=\bz= v\,\text{diag}(z_1,\ldots,
z_{N_c})$, but relax the requirement~(\ref{eq:constr_2}). For
instance, we may consider the following~\emph{perturbative} vacuum 
\be
\label{eq:pert_vac}
z_ k = z_k^{(0)}+ \eps(\xi) z_k^{(1)} + \eta (\xi) z_k^{(2)} + \ldots
\ , 
\ee
with $z_k^{(i)}$'s complex and subject to $\sum_k
z_k^{(i)}=0~\forall~i$, in order for the unimodularity
constraint~\eqref{eq:unimod_constr} to be satisfied. In the above,
$\eps(\xi)\sim c_1^\eps\xi^2 + c_2^\eps\xi^4 +\ldots,\eta(\xi)\sim
c_1^\eta \xi^2 +\ldots$~admit perturbative expansions in terms of the
coupling and can be determined iteratively at each order by plugging
$z_k$ into~(\ref{eq:1loop_CW}) and requiring that the equations be
satisfied.  As a proof of concept, let us pick the following specific,
but by no means unique, one-loop vacuum
\be
\label{eq:Zvev_pert_vac}
z_k= \bz_k = e^{2\pi i(k-1)/N_c} +c_1^\eps\xi^2\, e^{-2\pi i(k-1)/N_c}
\ , 
\ee
such that $\sum_k z_ k^2 =2c_1^\eps\xi^2\, N_c \neq 0$ and $\sum_k
z_k^2 \log z_k\approx N_c/2$. At order $\xi^2$, only the terms
proportional to~$\tr(z^2\log z)$~and~$\tr(\bz^2\log \bz)$~contribute
from the right-hand side of the equations~(\ref{eq:1loop_CW}). It is
straightforward to see that $c_1^\eps = \pm i/4$, meaning that, up to
a factor of $1/2$, $\eps(\xi)$ coincides with $\a_1^2$ at one loop
order, i.e. an acceptable one-loop flat direction is
\be
z_k=e^{2\pi i(k-1)/N_c}+\f{\a_1^2}{2}\, e^{-2\pi i(k-1)/N_c} \ .
\ee

Before we move to the discussion of multiloop contributions to the CW
potential, let us note en passant, that for massless excitations, the
one-loop contributions to the effective potential vanish
identically. This means that vacua for which the fields are either not
related by complex conjugation and only one of $z$, $\bz$ is
nonvanishing, or are nilpotent matrices, do not receive any
corrections at the one-loop level. Actually, this holds true at all
orders of perturbation theory as we will see in a while. This is due
to the chirality of the theory that allows for specific types of
vertices only, as well as the masslessness of the particles running in
the loops. Of course, such flat directions are in a sense quite
peculiar, as the CI is spontaneously broken but the spectrum of the
theory does not accommodate any massive particles, in the large-$N_c$
limit.

\subsubsection*{Higher-loop corrections to the effective potential}

\begin{figure}[!b]
  \centering
  \includegraphics[scale=.3]{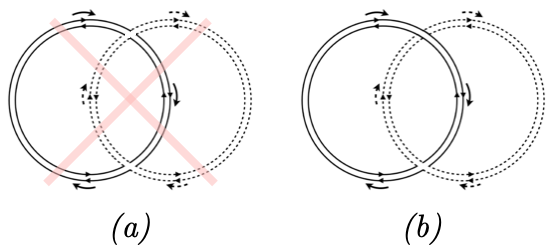}\hfill
  \caption{$(a)$ The chirality of the theory forbids some of the
    diagrams that would contribute to the effective potential at higher
    orders, such as the one above appearing in $\mathcal O ( \x ^
    4)$.~$(b)$ Example of a possible non-planar vacuum diagram in the
    leading $\xi^2$ order, to be neglected in the 't Hooft limit.}
  \label{fig:higher}
\end{figure}

\begin{figure}[!h]
    \centering
    \includegraphics[scale=.3]{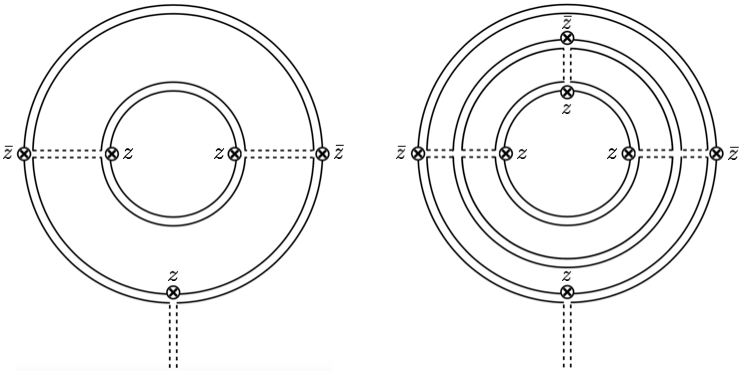}
    \caption{Exemplary multiloop tadpole diagrams of the fishnet type
      stemming from the single trace term. The specific three loop diagram
      (left) contains only cubic vertices, while the six loop one (right)
      contains both cubic as well as quartic vertices. The chirality of the
      theory forces the higher order planar diagrams stemming from the
      single-trace term to be of this form.}
    \label{fig:multiloop_tadpoles}
\end{figure}

Let us now proceed to the possible multiloop corrections to the
effective potential and study under which conditions and/or
modifications our considerations persist.

Let us focus first on the contributions from the single trace term of
the potential, $\Tr(\xoverline X\xoverline Z X Z)$.~When the
Lagrangian is expanded around the symmetry-breaking vacua, see
Appendix~\ref{sec:appendix}, the cubic and quartic terms give rise to
the ``chiral'' vertices presented in
Fig.~\ref{fig:vertices_full}\emph{(a)}. In essence, we may view the
trivalent vertices of the theory as the quartic chiral vertex with one
of the legs removed and replaced by the corresponding expectation
value, but otherwise preserving its double-line structure and
chirality. Their presence leads to planar graphs similar to the ones
built exclusively with a chiral quartic vertex, but with some
propagators, or parts of the closed loops of $X$ (or $Z$) propagators
removed (we call them loops with amputated propagators). It is
important to keep in mind that the non-Hermiticity of the FCFT
translates into a fixed chirality of the vertices. In other words, the
absence of the complex conjugate counterpart of $\Tr(\xoverline
X\xoverline Z X Z)$ is in one-to-one with the absence of the
``anti-chiral'' vertices presented in
Fig.~\ref{fig:vertices_full}\emph{(b)} and marked with red.

It is now straightforward to see that without the anti-chiral
vertices, the ``zoo'' of possible Feynman diagrams is rather
restricted.  For example, the diagram Fig.~\ref{fig:higher}$(a)$~with
two quartic vertices is absent from FCFT, even on top of vacua
breaking conformal symmetry, due to the opposite chirality of the
single-trace vertices there. Note that many more kinds of graphs exist,
like the one given in Fig.~\ref{fig:higher}$(b)$, with one quartic
vertex but with higher than spherical topology. They will certainly
modify the CW potential in the \(1/N^2_c\) order of the 't~Hooft
expansion, which we don't consider here. Of course this simplifies
considerably the situation. Nevertheless, the effective potential at
higher orders may receive contributions---among others---from fishnet
diagrams (with possibly amputated propagators, as explained
above). Two graphs of this type are presented in
Fig.~\ref{fig:multiloop_tadpoles}. Their types, and hence their
number, are very limited w.r.t. the generic graphs of scalar QFT at
each order in perturbation theory; unfortunately, they are still too
complicated for explicit computations~\footnote{We thank the referee
of the earlier version of this paper for pointing us on some of these
graphs.}.  For our purposes, however, it suffices to understand what
happens qualitatively. In the $z$-vacua under consideration, such
graphs can be only made of nested concentric circles of
$X$-propagators connected by "radial" $Z$-propagators (possibly
crossing the circles via quartic vertices) that end up on cubic
vertices. Note that the $X$ propagator (and the off-diagonal parts of
the $Z$ propagator) can connect cubic vertices of the same type only,
in contrast to the diagonal components of the $Z$ propagator that
necessarily connect different vertices.

\begin{figure}[!h]
    \centering
    \includegraphics[scale=.4]{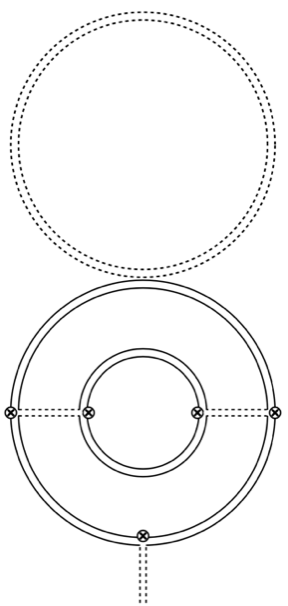}
    \caption{Typical planar graph that feeds into the (derivatives of
      the) effective potential at four loops. }
    \label{fig:DTsplit}
\end{figure}

As for the diagrams following from the double-trace terms, they can
only contribute to the large-$N_c$ limit if they occur in the graphs
in such a way that they connect two, otherwise disconnected, parts of
the graph (corresponding to each of two traces from the double-trace
vertex)~\cite{Gromov:2018hut,Fokken:2014soa}. An example of such graph
is drawn in Fig.~\ref{fig:DTsplit}. Like in the one-loop
considerations, the contributions coming from the double-trace terms
must exactly cancel the $\mu$-dependence from the single-trace terms,
as required by the conformality of the FCFT.

Several important comments are in order here. First of all, for the
``exotic'' vacua in which $\bz=0$ and $z\neq 0$ or vice-versa, the
tree-level potential is exact. In other words, it receives no loop
corrections, at any order in perturbation theory. This is either
because the diagrams cannot be constructed to start with, or they
vanish identically (in dimensional regularization) since the particles
running in the loops are massless.

The same is also true for the nilpotent vacua. Although both types of
vertices may be present (assuming that the fields are related by
complex conjugation), the corresponding diagrams vanish automatically,
either because they are proportional to traces of $z^p = 0$, for $p\ge
2$ (one always finds such traces for the innermost circle of
Fig.~\ref{fig:multiloop_tadpoles}), or because, again, the excitations
are massless.

One cannot conclude the same for the more interesting symmetry
breaking solutions of the previous section, when both $z,\bz$ are
diagonal. Then the higher order diagrams in the effective potential,
such as of the type presented on Fig.~\ref{fig:multiloop_tadpoles}, do
not vanish. To see this more clearly, let us focus on the three loop
graph on the left of this figure. On top of the diagonal
vacuum~(\ref{eq:Zvev}), it is equal to
\begin{equation*}
\begin{aligned}
\label{4-loopCW}
&\bar v v^2\l(\tilde\xi^{2}\r)^4\sum_{k,l,m}\bar z _ k^2 z_l^3 
\int\frac{ d^4p_1\,d^4p_2\,d^4p_3/(4\pi)^3}{p_1^4(p_2^2+\bz _ k z_
  j)((p_1+p_2)^2+z _k z_m)}\times \\ 
&~~~~~~~~~~~~~~~~~~\times\frac{1} {(p_3^2+\bz_m z
  _l)((p_1+p_3)^2+\bz_j z _ l)^2} \ , 
\end{aligned}
\end{equation*}
where $p_i$'s are dimensionless. This integral has logarithmic UV
divergences but once we add to it all diagrams of the same loop order
(containing double-trace vertices as well) it is guaranteed by
conformal symmetry that, as in the one-loop case, the overall result
will be nonzero, finite and scheme independent. Actually, it would be
interesting to explicitly compute it, a difficult but not impossible
problem, which however lies beyond our purposes here.

On general dimensional grounds, we expect loop corrections to the
effective potential to be of the following form
\begin{equation}
\label{eq:all-loop_CW}
V _ {n-loop} =c_{n} \l(\tilde \xi^2\r)^{n+1} \Tr (z ^ 2)\,
\Tr ({\zb} ^ 2)  f^{(n)}[z, \zb] \ ,
\end{equation}
where $c_{n}$ are numerical factors, and $f^{(n)}[z, \zb]$ homogeneous
functions of degree zero w.r.t. $z$ and $\bz$, symmetric w.r.t.~the
permutations of pairs of eigenvalues $(z_j,\bz_j)\to (z_k,\bz_k)$. For
instance, at the one loop level
\be
f^{(1)}  =\frac{\Tr \left(z^ 2 \log\frac{z}{\sqrt{Q}} \right)}{\Tr (z^
  2) }+   \frac{\Tr \left(\bar z^ 2 \log\frac{\bar z}{\sqrt{Q}}
  \right)}{\Tr (\bar z^ 2) } \ , 
\ee 
while at two loops, schematically 
\begin{equation*}
    \begin{aligned}
\label{2-loops}
f^{(2)} &\propto f^{(1)} +c\left(\frac{\Tr \left(z^ 2
\log^2\frac{z}{\sqrt{Q}} \right)}{\Tr (z^ 2) }+  \frac{\Tr
\left(\bar z^ 2 \log^2\frac{\bar z}{\sqrt{Q}} \right)}{\Tr (\bar z^ 2) }
\right)\ . 
\end{aligned} 
\end{equation*}

Like we did in the one-loop approximation, we have a number of
options. The first is to require that the higher-loop contributions
vanish independently from the ones coming from the lowest orders. This
would mean that in addition to~(\ref{eq:constr_2})
and~(\ref{eq:constr_3}), we have to further restrict the flat
directions, since we will encounter new patterns of traces in higher
loops. For example, for the 2-loop correction we will have to impose
$\Tr \left(z^ 2 \log^2z\right)=0$ as well. To fulfill simultaneously
all the flatness constraints, will certainly take larger than the
4$\times$4 sub-matrices we worked with previously. This is a procedure
that has to be effectuated repeatedly, and it is conceivable that more
than one conditions may be required at each loop order.

Alternatively, we may insist that the tree level and one-loop
contributions vanish independently from each other by virtue
of~(\ref{eq:constr_2}) and~(\ref{eq:constr_3}), while the higher loop
corrections are taken care of by ``perturbing'' this solution in the
sense of~(\ref{eq:pert_vac}). This way, all quantum corrections
starting from a specific loop order will be comparable by design so
they may compensate for each other.

Finally, we can stick with the perturbative vacua~(\ref{eq:pert_vac})
and appropriately generalize them by keeping higher powers of $\xi$
and even use different harmonics so that all the terms in the
effective potential will be of the same order. By doing so, we need
only to impose one condition per loop order: that is, the
derivative(s) of the full effective potential w.r.t. the fields be
zero. This option is attractive since, in principle, we have the
possibility to study it perturbatively, order by order.

\section*{Conclusions and open problems}

In this work we initiated the study of spontaneous conformal symmetry
breaking in the recently proposed fishnet CFT.  We showed that the
theory admits a plethora of classical flat directions along which
conformal symmetry is nonlinearly realized without fine-tuning.  We
also studied the quantum corrections and found that the classical
conformal invariance is not violated, at least in some subclasses of
the classical solutions. This fact is the (nontrivial) aftermath of a
delicate interplay between the finiteness of the theory, its
non-Hermiticity, the large-$N_c$ limit and the constraints on the flat
directions.

The FCFT is integrable in 't~Hooft limit. Although the integrability
is demonstrated only for the unbroken vacuum, some features of it may
survive for the broken vacua, at least in perturbation theory. This
could offer a unique opportunity to elucidate various aspects of the
dynamics behind spontaneous symmetry breaking in this particular
theory. At the same time, it can serve as an inspiring example for
CFTs with such behavior in general. A first step towards this
direction could be to check the validity of the constraints that were
derived in~\cite{Karananas:2017zrg}. For instance, the deep infrared
limit of the two-point functions of scalar primary operators $\mathcal
O _ I$ were shown to obey the identity
\be
\label{eq:cons_cond_corr}
\av{\mathcal O _ I} \av{\mathcal O _ J } \sim \lim _ {x\to \infty}
\sum _ K \f {c _ {IJK}} {|x|^ {\D _ I +\D _ J -\D _ K} }\av {\mathcal
  O _ K}\ , 
\ee
with $c_ {IJK}$ the OPE coefficients and $\D$'s the corresponding
scaling dimensions. As a test of this relation in the context of the
FCFT, we can consider the dimension-two operators
$
\Tr(X Z),~\Tr(X\xoverline Z),~
\Tr(\xoverline X Z),~\Tr(\xoverline X \xoverline Z),
$
whose two-point correlators in the planar limit are protected against
quantum corrections and decay as $\sim |x| ^
{-4}$~\cite{Gromov:2018hut}.

The fact that the vev of these operators vanish for our vacua,
immediately implies the validity of~(\ref{eq:cons_cond_corr}). The OPE
data for these operators in the unbroken vacuum have been computed
in~\cite{Gromov:2018hut}. A more detailed study of various consistency
conditions is left for future work.

In particular, the scalar one-point functions of the operators
entering the r.h.s. of these operators might be computable, using the
methods developed in~\cite{Grabner:2017pgm,Gromov:2018hut}.

Let us also point out that some of the (classical) vacua we discussed
in this work are present in the full $\g$-deformed $\mathcal N =
4$~SYM and propagate all the way to the fishnet CFT. One can, for
instance, assume that $x _ {\rm tree} = c\, z _ {\rm tree}$, with $c$
a constant. Requiring that the above satisfy the equations of motion
of the $\g$-deformed theory even before the fishnet double scaling
limit is taken, translates into the coefficient $\alpha_2$ of the
double-trace terms involving both $Z$ and $X$
in~(\ref{eq:double-trace}) being completely fixed $ \alpha _ 2 = -4 g
^ 2 \sin ^ 2 \l(\f{\g _ 3}{2}\r) \ .  $ As a sanity check, note that $
\displaystyle \lim_{\substack{g\to 0 \\ \g_3\to i\infty}}\alpha _ 2
\sim \x ^ 2 \ ,~~~\text{while}~~~\lim _ {\g_3\to 0}\alpha _ 2 \sim 0 $
as it should. To put it differently, the mere requirement that the
full $\g$-deformed ${\cal N}=4$ SYM theory possesses flat directions
is smoothly connected to the ones of its fishnet ``descendants,''
completely determines one of the coefficients appearing in the action
of the full original CFT.

Finally, it would be interesting to study to what extent the discussed
properties of the FCFT survive in the next $1/N_c$ orders, or even for
finite $N_c$.

\leavevmode

\begin{acknowledgments}
We thank B. Basso, G. Korchemsky, A. Zhiboedov and D. Zhong, for
useful discussions and comments. We are indebted to the anonymous
referee for insightful and constructive comments that significantly
improved the manuscript.~G.K.K. would like to thank CERN and EPFL for
the warm hospitality during the first and last stages of this
project. The work of G.K.K. was partially funded by the Deutsche
Forschungsgemeinschaft (DFG, German Research Foundation) under
Germany's Excellence Strategy EXC--2111--390814868.  V.K. is grateful
to CERN Theory Division for the kind hospitality and support during
his CERN association term. The work of M.S. was supported by the
ERC-AdG-2015 grant 694896 and the Swiss National Science Foundation.
\end{acknowledgments}

\bibliographystyle{utphys}
\bibliography{BrokenCFT}
\onecolumngrid
\appendix
\section{}
\label{sec:appendix}
Once we shift the fields as in~(\ref{eq:shifts}), the relevant parts
of the Lagrangian for the excitations read
\be
\label{eq:app1}
\begin{aligned}
&\mathscr L'=\mathscr L +\mathscr L_{d.t.} +\mathscr L_{(2)}+\mathscr
L_{(3)} \ ,
\end{aligned}
\ee
where $\mathscr L$ and $\mathscr L_{d.t.}$ have the same form as
in~(\ref{eq:lagr_biscalar})~and~(\ref{eq:double-trace}), while
\be
\label{eq:app2}
\begin{aligned}
 &\mathscr  L_{(2)}/(4\pi)^2=N_c\,\xi^2 \Tr\, \Big[ \bx \bz X Z + \bx
 \bZ x Z+\bx\bZ X z+\bX \bz x Z+ \bX \bz X z +\bX \bZ xz\Big]\\ 
&+\alpha_ 1 ^ 2  \Big[
\Tr(x^2)\Tr(\bX^2)+\Tr(X^2)\Tr(\bx^2)+4\Tr(xX)\Tr(\bX\bx)+\Tr(z^2)\Tr(\bZ
^ 2)+\Tr(Z^2)\Tr(\bz^2)+4\Tr(z Z)\Tr(\bZ\bz) \Big]\\ 
&-\alpha_2^2  \Big[ \Tr(xz)\Tr(\bX \bZ) +
\Tr(XZ)\Tr(\bx \bz) +\Tr(xZ)\Tr(\bx \bZ) +\Tr(Xz)\Tr(\bX \bz)
+\Tr(xZ)\Tr(\bX \bz) + 
\Tr(Xz)\Tr(\bx \bZ)\Big] 
\\&~~~~~~~+ \Tr(x\bz)\Tr(\bX Z) +
\Tr(X\bZ)\Tr(\bx z) +\Tr(x\bZ)\Tr(\bx Z) +\Tr(X\bz)\Tr(\bX z)
+\Tr(x\bZ)\Tr(\bX z) +\Tr(X\bz)\Tr(\bx Z)\Big] \ , 
\end{aligned}
\ee
and 
\be
\label{eq:app3}
\begin{aligned}
&\mathscr L_{(3)}/(4\pi)^2 =N_c\,\xi^2\Tr\Big[\bx\bZ X Z+ \bX \bz X Z+
\bX \bZ x Z +   \bX \bZ X z  \Big]\\ 
&+2\alpha_ 1 ^ 2  \Big[
\Tr(Xx)\Tr(\bX^2)+\Tr(X^2)\Tr(\bX\bx)+\Tr(Zz)\Tr(\bZ ^
2)+\Tr(Z^2)\Tr(\bZ \bz) \Big] 
\\
&-\alpha_2^2  \Big[ \Tr(xZ)\Tr(\bX \bZ) +
\Tr(Xz)\Tr(\bX \bZ) +\Tr(XZ)\Tr(\bx \bZ) +\Tr(XZ)\Tr(\bX \bz) 
\\
&~~~~~~~+\Tr(x\bZ)\Tr(\bX Z) +\Tr(X\bz)\Tr(\bX Z)+\Tr(X\bZ)\Tr(\bx Z)
+\Tr(X\bZ)\Tr(\bX z)  \Big] \ . 
\end{aligned}
\ee
Using~(\ref{eq:app2}), we can read the quadratic forms for the
excitations $X$ and $Z$ of the fields at the large-$N_c$ limit. Moving
to momentum space, on top of the $z$ flat directions $x=\bx=0$, we
find
\be
\label{eq:app_mass_matr}
\mathcal D_{X}^{-1}=N_c\begin{pmatrix}
k^2+\tilde \xi^2 \,\bz\times z & 0 \\
0 & k^2 +\tilde \xi^2\,  z\times\bz 
\end{pmatrix} \ ,~~~
\mathcal D_{Z}^{-1}=\begin{pmatrix}
N_c\, k^2 & 2(4\pi\a_1)^2\Tr(z^2)\\
2(4\pi\a_1)^2\Tr(\bz^2) & N_c\,k^2
\end{pmatrix} \ ,
\ee
where $\times$ denotes the matrix product. The masses of the
excitations can be easily found from the above by setting $k^2=0$. For
the $X$ and $\bX$, these correspond to the eigenvalues of the matrices
$\tilde \xi^2 \bz\times z$ and $\tilde \xi^2 z\times\bz$, while for
$Z$ and $\bZ$, the masses are $\pm 2(4\pi\a_1)^2 \sqrt{Q}$. In turn,
their exact values depend on the choice of the flat directions. For
instance, if we move along~(\ref{eq:Zvev})-(\ref{eq:constr_2}), the
$X$'s masses are proportional to $|v|^2 \tilde \xi^2 \bz_i z_j$, while
the $Z$'s are massless. On the other hand, for the nilpotent matrices
or the configurations with $\bz = 0$ while $z\neq 0$, the spectrum of
the theory comprises only massless excitations, since the eigenvalues
of both $\mathcal D_X^{-1}\Big\vert_{k^2\to 0}$ and $\mathcal
D_Z^{-1}\Big\vert_{k^2\to 0}$ are zero.

Inverting~(\ref{eq:app_mass_matr}), we find the corresponding
propagator matrices
\bea
\label{eq:app_prop_matr}
\mathcal D_{X}=\f{1}{N_c}\begin{pmatrix}
\frac{1}{k^2+\tilde \xi^2 \,\bz\times z} & 0 \\
0 & \frac{1}{k^2 +\tilde \xi^2\,  z\times\bz} 
\end{pmatrix},~
\mathcal D_{Z}=\frac{-1}{N_c^2k^4-4(4\pi\a_1)^4\Tr(z^2)\Tr(\bz^2) 
}&\begin{pmatrix} 
-N_c k^2 & 2(4\pi\a_1)^2\Tr(z^2)\\
2(4\pi\a_1)^2\Tr(\bz^2) & -N_c  k^2 
\end{pmatrix}, &
\eea
with which we can immediately compute loops. From~(\ref{eq:app1})
and~(\ref{eq:app3}), we can deduce the Feynman rules for the theory;
some of the vertices associated with the single-trace terms are
presented in Fig.~\ref{fig:vertices_full}.

\section{}
\label{sec:appendix2}

To be maximally pedagogic, let us study in some details the one-loop
diagrams appearing in Fig.~\ref{fig:1loop_tadpoles}, for general
diagonal flat directions. Let us focus on the
graph~\ref{fig:1loop_tadpoles}$(a)$ coming from the single-trace term
with an insertion of the vev $\bz$~\footnote{This means that we are
actually computing the derivative of the one-loop correction
w.r.t. $z$.}. Reading the corresponding vertex from~(\ref{eq:app2})
and using $\mathcal D_X$ from~(\ref{eq:app_prop_matr}), we find that
the diagram evaluates to
\be
\includegraphics[valign = c,scale=.5]{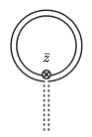}=
\displaystyle\bar v\tilde \xi ^ 2 \sum_i \bar z _ i\int
\f{d^4k}{(2\pi)^4}\frac{1}{k^2 +  \tilde \xi ^ 2\bar v v\bar z _ i z _
  j}=\bar v^2 v\l(4\pi\xi^2\r)^2 z_j \sum _ {i}  
\bar z _ i^2 \l[-\f {1} {\bar \eps}+\log \l(\f{\tilde \xi ^2 \bar v v
}{\m^2}\r)+\log \l(\bar z _ i z _j\r)\r ] \ , 
\ee
 where we introduced $1/\bar \eps = 1/\eps+1 -i\pi-\gamma_E
+\log(4\pi)$, $\g_E\approx 0.5772$ is the Euler-Mascheroni constant
and $\m$ is the renormalization scale.

For the ``compensating'' double-trace
diagram~\ref{fig:1loop_tadpoles}$(b)$, we should look at
eq.~(\ref{eq:app3}) and work with the 21-component of the $\mathcal
D_Z$ propagator; we obtain
\be
\begin{aligned}
\includegraphics[valign = c,scale=.5]{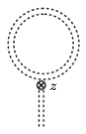}
&=-\bar v^ 2 v \,4(4\pi\a_1)^4 z_j \sum_i \bz_i^2\bigintsss
\f{d^4k}{(2\pi)^4}\frac{1}{\displaystyle k^4-4(4\pi\a_1)^4(\bar v
  v)^2\sum_{l,m}\bz_l^2
  z_m^2}\phantom{~~~~~~~~~~~~~~~~~~~~~~~~~~~~~~~~~~~~}\\[-5pt] 
&= \bar v^2 v \f{(4\pi\a_1)^4}{4\pi^2}z_j \sum_i \bz_i^2 \l[-\f {1}
{\bar \eps}+\log \l(\f{-2i(4\pi\a_1)^2\bar v v }{\m^2}\r)+\f 1 2\log
\sum_{l,m}\bz_l^2 z_m^2\r ] \ . 
\end{aligned}
\ee

Putting the two contributions together and using the one-loop value
$\a_1^2=\pm i \xi^2/2$, it is straightforward to see that the $1/\bar
\eps$ piece as well as the logarithms containing $\m$ cancel
automatically, as it should be in the CFT. Switching back to matrix
notation, the derivative of the one-loop contribution w.r.t. $z$ reads
\be
\frac{\partial }{\partial z}V_{1-loop} = \f{\tilde \xi^4}{(4\pi)^2} z
\l[ \Tr(\bz^2)\log\frac{z}{\sqrt{Q}} +\Tr\left( \bz^2\log\frac{ \bz
  }{\sqrt{Q}}\right)\r] \ , 
\ee
where $Q =\sqrt{\tr(z^2)\,\tr(\bz^2)}$ was also defined under
eq.~(\ref{eq:1loop_CW}). Integrating the above over $z$, we readily
obtain
\be
\label{eq:app_one_loop_eff}
V_{1-loop}= \f{\tilde \xi^4}{32\pi^2}\l[\Tr(z^2)\Tr\l(\bz^2\log
\f{\bz}{\sqrt{Q}}\r)+\Tr(\bz^2)\Tr\l(z^2\log \f{z}{\sqrt{Q}}\r)\r] \
. 
\ee
Following exactly the same steps for the conjugated diagrams, we
obtain the derivative of the one-loop contribution w.r.t. $\bz$
\be
\frac{\partial }{\partial \bz}V_{1-loop} = \f{\tilde \xi^4}{(4\pi)^2}
\bz \l[ \Tr(z^2)\log\frac{\bz}{\sqrt{Q}} +\Tr\left( z^2\log\frac{ z
  }{\sqrt{Q}}\right)\r] \ . 
\ee

\end{document}